\documentstyle{elsart}
\input{psfig}
\begin{document}

\begin{frontmatter}

\title{Search for Exotic Strange Quark Matter in High Energy Nuclear Reactions}

\author[pennst]{T.A. Armstrong},
\author[yale]{K.N. Barish\thanksref{newken}},
\author[wayne]{S.J. Bennett},
\author[yale]{A. Chikanian},
\author[yale]{S.D. Coe},
\author[wayne]{T.M. Cormier},
\author[purdue]{R. Davies},
\author[bari]{G.De Cataldo},
\author[wayne]{P. Dee},
\author[yale]{G.E. Diebold},
\author[brookhaven]{C.B. Dover\thanksref{newdover}},
\author[wayne]{P. Fachini},
\author[yale]{L.E.Finch},
\author[yale]{N.K. George},
\author[bari]{N. Giglietto},
\author[vanderbilt]{S.V. Greene},
\author[mit]{P. Haridas},
\author[iowa]{J.C. Hill},
\author[purdue]{A.S. Hirsch},
\author[iowa]{R. Hoversten},
\author[ucla]{H.Z. Huang},
\author[wayne]{B. Kim},
\author[yale]{B.S. Kumar},
\author[westpoint]{T. Lainis},
\author[yale]{J.G. Lajoie},
\author[pennst]{R.A. Lewis},
\author[iowa]{B. Libby\thanksref{newbruce}},
\author[yale]{R.D. Majka},
\author[wayne]{M.G. Munhoz},
\author[yale]{J.L. Nagle\thanksref{newnagle}},
\author[mit]{I.A. Pless},
\author[yale]{J.K. Pope},
\author[purdue]{N.T. Porile},
\author[wayne]{C.A. Pruneau},
\author[umass]{M.S.Z. Rabin},
\author[bari]{A. Raino},
\author[pennst]{J.D. Reid\thanksref{newreid}},
\author[purdue]{A. Rimai\thanksref{newrimai}},
\author[yale]{F.S. Rotondo},
\author[yale]{J. Sandweiss},
\author[purdue]{R.P. Scharenberg},
\author[yale]{A.J. Slaughter},
\author[pennst]{G.A. Smith},
\author[bari]{P. Spinelli},
\author[purdue]{M.L. Tincknell},
\author[pennst]{W.S. Toothacker},
\author[mit]{G. Van Buren},
\author[iowa]{F.K. Wohn},
\author[yale]{Z. Xu},
\author[wayne]{K. Zhao}

\centerline{\it{(The E864 Collaboration)}}

\address[bari]{University of Bari/INFN, Bari, Italy}
\address[brookhaven]{Brookhaven National Laboratory, Upton, New York 11973}
\address[ucla]{University of California at Los Angeles, Los Angeles, California 90024}
\address[iowa]{Iowa State University, Ames, Iowa 50011}
\address[umass]{University of Massachusetts, Amherst, Massachusetts 01003}
\address[mit]{Massachusetts Institute of Technology, Cambridge, Massachusetts 02139}
\address[pennst]{Pennsylvania State University, University Park, Pennsylvania 16802}
\address[purdue]{Purdue University, West Lafayette, Indiana 47907}
\address[westpoint]{United States Military Academy, West Point, New York 10996}
\address[vanderbilt]{Vanderbilt University, Nashville, Tennessee 37235}
\address[wayne]{Wayne State University, Detroit, Michigan 48201}
\address[yale]{Yale University, New Haven, Connecticut 06520}

\thanks[newken]{Present address: University of California at Los Angeles, Los Angeles, CA 90024}
\thanks[newdover]{Deceased.}
\thanks[newbruce]{Present address: Department of Radiation Oncology, Medical College of Virginia, Richmond, VA 23298}
\thanks[newnagle]{Present address:  Columbia University, Nevis Laboratories, Irvington, NY 10533}
\thanks[newreid]{Present address: Vanderbilt University, Nashville, TN 37235}
\thanks[newrimai]{Present address: Institut de Physique Nucl\'{e}aire 91406 ORSAY Cedex, France}

\begin{abstract}
We report on a search for metastable positively and 
negatively charged states of strange quark matter 
in Au + Pb reactions at 11.6 A GeV/c in experiment E864.
We have sampled approximately six billion 10\% most central Au~+~Pb
interactions and have observed no strangelet states 
(baryon number A $<$ 100 droplets of strange quark matter).  
We thus set upper limits on the production 
of these exotic states at the level of $1-6 \times 10^{-8}$ per
central collision.  These limits are the best and most model independent for this
colliding system.  We discuss the implications of our results on
strangelet production mechanisms, and also on the stability 
question of strange quark matter.
\end{abstract}

\end{frontmatter}

\section{Introduction}

All observed color singlet states involve either three quarks (baryons) or
quark-antiquark pairs (mesons).  However, the Standard Model which describes
these states does not forbid the existence of
color singlet states of more than three quarks (for example a bag of 18
quarks).  It is known that such quark matter states made from only up and 
down quarks are less stable than normal nuclei of the same 
baryon number A and charge Z, since
nuclei do not decay into quark matter.  
However, if such objects were made of three quark flavors, including
strange quarks, they could gain stability from a reduction 
in the Fermi energy, despite the additional mass of the strange quark.  
Present theoretical understanding
of strange quark matter states indicates that they are potentially
metastable \cite{Farhi}, and even possibly more stable than 
nuclear matter \cite{Witten}.  Due to the lack
of theoretical constraints on bag model parameters and 
difficulties in calculating color magnetic interactions 
and finite size effects \cite{Schaffner}, 
the issue of the stability of strange quark matter is 
presently an experimental one.

There have been searches for strange quark matter in terrestrial matter \cite{Hemmick}, in
cosmic rays and in astrophysical objects \cite{Alcock} 
(for a review see \cite{Kumar}).  However, the most controlled investigation which
attempts both production and detection to date is 
the search for strangelets in relativistic
heavy ion collisions.  Heavy ion collisions are a good environment to look
for such states for three reasons.  First, high energy heavy ion reactions are
the only colliding system in the laboratory which produce 
significant strangeness and baryon number in a small volume from which a
strangelet might be formed.  Second, it is believed that in these collisions
a phase transition to a quark-gluon plasma might occur.  If a heavy ion collision
results in the formation of hot quark matter it might then cool into
a metastable state of cold strange quark matter.  And last,
because the accelerator is under the experimenters' control, we can study a
large number of reactions and set low sensitivity levels in the absence of 
observation.

\section{Experiment}

Experiment 864 at the Brookhaven AGS facility was specifically designed for
the purpose of searching for strangelets.  Strangelets are expected to
have a unique experimental signature of a low charge to mass ratio 
(below the range of known nuclear
isotopes) due to the roughly equal numbers of up, down and strange quarks (charge +2/3,
$-$1/3, and $-$1/3).  The experiment identifies secondary particles produced in Au + Pb
collisions at 11.6 GeV/c per nucleon by measuring their masses and charges.  The
experiment has a large geometric acceptance and can operate at high interaction
rates up to $10^{6}$ collisions per one second beam spill.  These two features 
allow the experiment to achieve a high level of sensitivity.

\subsection{Apparatus}

\begin{figure}
\centerline{\hbox{\psfig{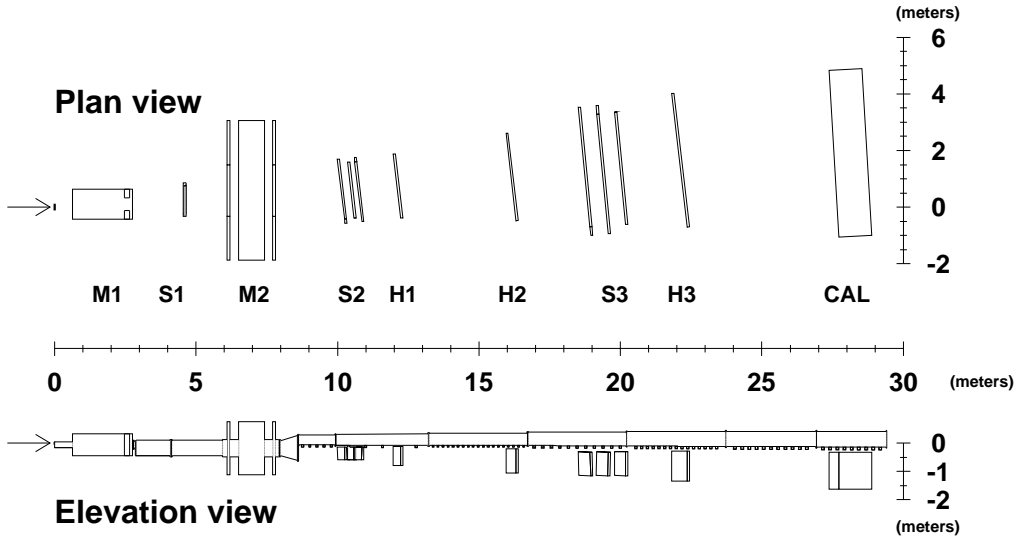}}}
\caption{Experiment 864 detector layout in plan and elevation view.  
The incident beam on the Pb target is shown as an arrow.  The two dipole 
magnets are labeled as M1 and M2.  Also shown are the straw stations (S1, S2, S3),
the three time-of-flight hodoscopes (H1, H2, H3) and the hadronic calorimeter (CAL).
One can see the full extent of the vacuum chamber in the elevation view.}
\label{fig:fig_detector}
\end{figure}

The apparatus is shown in Fig. \ref{fig:fig_detector}.
A collimated beam of Au ions (shown as an arrow) passes through a quartz plate Cerenkov
counter.  This beam counter is able to deliver timing information for each ion with a 
resolution of $\sigma_{t} \approx 100$~ps at incident rates up to $10^{7}$ ions
per one second spill.  This information is used as the starting time our velocity
measurements.  There are veto counters to reject beam particles 
outside our profile, upstream beam interactions, and events with two or more incident Au ions.  The 
target is a Pb disk and 30\% of an interaction length for Au.  
Non-interacting Au ions and beam fragments with low transverse momentum are contained in an
aluminum and steel vacuum chamber (shown in the elevation view) 
to reduce interactions in air which might shower secondary particles 
into the downstream spectrometer.

Downstream of the target are two dipole magnets (M1 and M2).  
The straw tube tracking detector (S1) inside the vacuum
chamber between the two magnets was not used in the analysis 
presented here.  Au ions which interact in the target produce secondary 
particles which, after passing through the two 
magnets, may exit the vacuum chamber through a thin window.  These secondaries
then traverse the remainder of the spectrometer below the vacuum chamber.  
Particles must have a downward vertical angle of at least $-$17.5 milliradians
to exit through the window.  

Those secondary particles within the acceptance
are tracked using two stations of straw tubes (S2 and S3) and three time-of-flight
hodoscopes (H1, H2 and H3).  Each of the time-of-flight hodoscopes 
consists of 206 plastic scintillator slats read out by two photo-multiplier
tubes (one at the top and one at the bottom).  The scintillators have 45 degree
diamond mill finished ends to reflect the light from the scintillator through
a 90 degree bend into a cylindrical lucite light guide.  This 90 degree bend
allows the detector to be placed as close to the vacuum chamber as possible,
thus extending our acceptance closer to zero degrees in the vertical direction.  The hodoscopes
give three independent measurements of the particle charge via energy loss $dE/dx$.  
Also each plane yields time-of-flight from the mean time of the two 
photo-multiplier tubes with a resolution of $\sigma_{t} \approx$ 120-150 ps.
The straw tube detectors are used to improve a track's spatial
resolution.  
%This improved pointing back into the magnetic field region results in a 
%momentum resolution of $\Delta p/p \approx$ 3\%, which is 
%limited by multiple scattering in the spectrometer material.  
Each straw tube station is made up of three planes ($x, u, v$), with one
oriented in the 
vertical and the others at $\pm$ 20 degrees relative to vertical.  The straws are 4 mm diameter
tubes and are stacked in a doublet configuration for each plane,
eliminating possible gaps in the detector.  

At the end of the spectrometer is a spaghetti design
hadronic calorimeter.  The calorimeter consists of 754 towers with
dimension $\rm{10~cm \times 10~cm \times 117~cm}$.  The calorimeter is
approximately five hadronic interaction lengths deep.  Each tower is constructed 
from grooved lead sheets with scintillating fibers approximately collinear
to the incident particle trajectories.  The ratio of lead to fiber is
chosen to approximately compensate the calorimeter energy
responses for hadronic and electro-magnetic showers.  This fiber
configuration leads to an excellent timing resolution $\sigma_{t} \approx 400$ ps
for hadronic showers.
The energy resolution for hadrons has been determined to be
 6\%~+~34\%/$\sqrt{E}$, where
$E$ is the energy deposited in units of GeV \cite{Nagle}.  More details on the calorimeter
performance are given in Ref \cite{NIM}.  

\subsection{Trigger}

The data acquisition system can record approximately 1500 events per one
second spill over a four second duty cycle, and thus in order to sample 
the interaction rate we have implemented
two triggers.  
The probability for strangelet production is expected to be significantly
increased in the most central collisions \cite{Baltz}.
Thus, a low level trigger selects approximately the 10\% most central (low
impact parameter) Au + Pb interactions.  This selection is made using a four
fold segmented scintillation counter covering forward angles 
from 16.6 to 45.0 degrees \cite{Haridas}.  
Interactions producing pulse heights in the 10\% highest fraction are selected.

\begin{figure}
\centerline{\hbox{\psfig{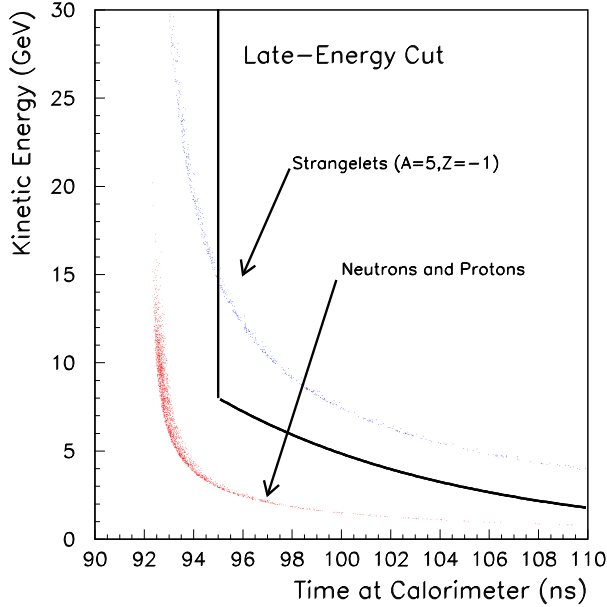}}}
\caption{Results from Monte Carlo simulation of the kinetic
energy versus the time-of-flight at the front face of the calorimeter
are shown for nucleons (neutrons and protons) and for strangelets (baryon
number A=5 and charge Z=$-$1).  The points assume perfect timing and
energy information.  Data from the experiment have significant
blurring of the distributions due to limited energy sampling and 
detector resolutions.  An example high mass trigger cut is shown
as a solid line.}
\label{fig:fig_let}
\end{figure}

The second trigger is a high mass trigger which is used to select events with
possible strangelet candidates and reject events with only normal hadrons in
our spectrometer.   The calorimeter measures the particles' kinetic energy
and time-of-flight (which is easily related to the velocity $\beta$).  We have
constructed a trigger connected to each calorimeter tower 
which has a programmable look-up table of accept values.  
Shown in Fig. \ref{fig:fig_let} is
a Monte Carlo generated kinetic energy and time-of-flight
distribution at the front face of the calorimeter (without detector
effects).  The upper band is 
simulated strangelets with A = 5 and Z = $-$1.  The lower band is simulated
neutrons and protons striking the calorimeter.  An example of a trigger curve
is shown in the figure.  Any event in which at least one tower measures an
energy and time-of-flight above the curve would be recorded.  
There is a minimum time-of-flight cut component to the trigger such that we only
accept candidates with rapidity approximately $y < 2.2$.  In the 
experiment, the distributions are significantly blurred by finite time
resolution, single tower energy sampling, and energy resolution.  However, we
were able to program the trigger such that only 2\% of ordinary central interactions
fire the high mass trigger, while maintaining greater than 90\% trigger
efficiency for strangelets of mass greater than 5 GeV/c$^2$.

\subsection{Data Sets}

The results discussed in this paper are from data taken in the fall of 1995.
For these studies we operated the spectrometer at two different magnetic field
configurations.  We recorded approximately 85 million events at what we refer
to as the ``$-$0.75T'' field setting which optimizes the acceptance for
negatively charged strangelet states while sweeping out copiously produced
hadrons (kaons, protons, etc.).  We also took approximately 27 million
events at the ``+1.50T'' field setting which optimizes the acceptance for
positively charged strangelet states.   Each data set was taken using both
the centrality trigger requirement and the high mass trigger.  
We have searched only for positively charged states in the ``+1.50T'' data sample;
however, in the ``$-$0.75T'' data, due to the larger number of events and the
lower magnitude of the field, there is significant sensitivity to positively 
as well as negatively charged states.  We combine the final results from each data 
sample to calculate our experimental limits.

\section{Analysis}

In this section, we present the details of the strangelet search
analysis.   A complete description of the analysis of the ``$-$0.75T'' data set 
may be found in \cite{Nagle}.    There are some differences in the analysis
of the ``+1.50T'' data set due to differences in detector occupancies
and background sources, not detailed in this paper.  Complete details of the
``+1.50T'' data set may be found in \cite{Coe}.

\subsection{Tracking}

We identify charged particle tracks
which have consistent hits in the straw tube chambers and the three time-of-flight
hodoscopes.  After all hits have been associated with a given track, we make
a global fit using all hit information to determine the particle's trajectory.  
From the trajectory of the track in the bend plane of the
magnetic field, we determine the particle's rigidity ($R = p /Z$) assuming the
track originated at our production target.    The charge of the particle is
measured from energy deposition ($dE/dx$) in each of the three
hodoscope planes.  The velocity $\beta$ is calculated from the path length
of the track and hodoscope time-of-flight 
information.  The mass of the track can now be calculated.
\begin{equation}
m = {{R \times Z} \over {\gamma \beta}}
\end{equation}  
Thus, for each track candidate we have a measure of the mass and charge.  
At the first stage in analysis, we save all events with 
possible strangelet candidates in the rapidity range $1.0 < y < 2.2$
with mass greater than 5 GeV/c$^2$ and any charge Z=$\pm$1,$\pm$2.  
The only known particles expected within this sample are $^{6}$He and $^{8}$He isotopes.  
We find that only a small percentage (1-2\%) of the recorded events contain candidates.

In doing a sensitive search for new particle states, it is critical to have 
significant redundancy in various measurements in order to reduce possible
background which might yield false signals for strangelets.  The first level
of redundancy is in the tracking system described above.  Through multiple 
tracking measurements, we over determine the particle trajectory in space and time.
We have fit to the track velocity and horizontal and vertical path, and
the $\chi^{2}$ value from each fit is examined.  Known hadronic species measured in the
same data sample (antiprotons, deuterons, $^{3}$He isotopes) show
good agreement with the theoretical $\chi^{2}$ calculations for the correct number of
degrees of freedom.  This agreement gives us confidence that we understand
the alignments, calibrations and resolutions of the tracking detectors.  We apply
track quality cuts on these $\chi^{2}$ values with known efficiencies for ``good''
particles to further reject possible background tracks reconstructing as
strangelets.  The $\chi^{2}$ cuts are determined to be ~80\% efficient (as
verified with identified antiprotons).  When the cuts are applied to the set of strangelet
candidates, only approximately 20\% of the candidates remain.  This low percentage
indicates that the majority of the candidates are background, and thus
have significantly worse track quality.  

\begin{figure}
\centerline{\hbox{\psfig{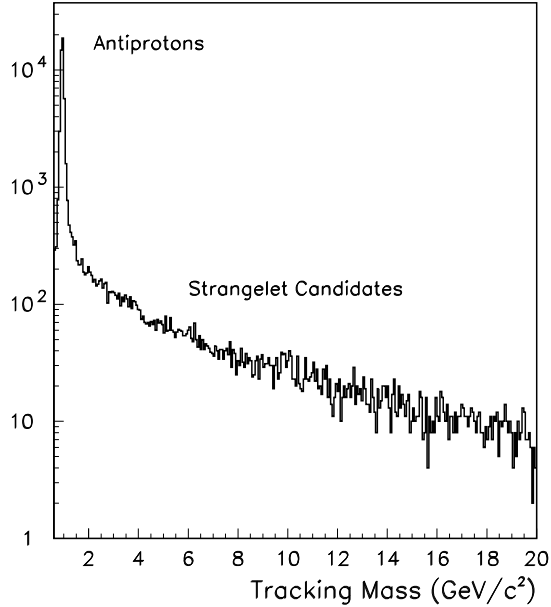}}}
\caption{The track reconstructed mass distribution for Z=$-$1 candidates
from the full ``$-$0.75T'' data sample.}
\label{fig:fig_cand_mass}
\end{figure}

From the ``$-$0.75T'' data sample, the 
distribution of reconstructed masses from tracking for Z=$-$1 candidates
surviving track quality cuts is shown in Fig. \ref{fig:fig_cand_mass}.
A prominent peak containing approximately 50,000
antiprotons is seen with an exponential tail of 
higher mass candidates.  The antiproton mass is at the Particle Data Book
value with a mass resolution of approximately $\Delta m/m \approx 4$\%, as
expected from our time-of-flight and momentum resolutions.
There are approximately 20,000 remaining Z~=~$-$1 candidates 
with mass greater than 5~GeV/$c^2$ (as shown in the figure).
Also, there are a few hundred Z~=~+1 candidates with mass greater than 5~GeV/$c^2$
from the ``+1.50T'' data sample.
There are no Z~=~$-$2 candidates and only a few Z~=~+2 candidates with
masses beyond the mass for $^6$He. 

\subsection{Background Sources}

There is a known source of background in our spectrometer which can
create high mass candidates from track reconstruction.  A neutron leaving
the target can pass undeflected through a portion or all of the magnetic
field region and then undergo an inelastic reaction in the vacuum exit
window, the straw chamber in vacuum (S1), air, etc., emitting a
forward going proton.  The proton creates a charged particle track
in all of the downstream tracking detectors.  The track may have a reasonable
velocity, $\beta$, but the reconstructed rigidity can be large, since
the neutron trajectory did not bend in the magnetic field.   These erroneous
high rigidity measurements can yield high mass candidates.  We have done extensive
Monte Carlo simulations of this background reaction \cite{Barish}.  Despite the fact
that the resulting particle in the spectrometer is a proton, the track
can also mimic a Z=$-$1 particle since the track results from a neutron (whose
trajectory is unaffected by the magnetic fields).

This background process results only in Z=$\pm$1 candidates, which explains
the significantly lower number of Z=$\pm$2 candidates.  Also, there will
be some inconsistency in the tracking fits for these candidates, since the 
inelastic interaction has some kink angle between the incoming neutron
and outgoing proton.  This feature explains the low acceptance for the
$\chi^{2}$ cut for these candidates.  
However, for some of these background reactions
the track quality appears good and the tracking system is fooled.

\subsection{Hadronic Calorimeter}

The calorimeter makes an second measurement of the particle mass
completely independent of the magnetic field region:  
\begin{equation}
m = {{KE} \over {\gamma -1}}
\end{equation}
where KE is the kinetic energy of the particle and $\gamma$ is the relativistic
factor.
The tracks resulting from the background process described above should reconstruct with the 
mass of a proton in the calorimeter.  If the candidates are real strangelets,
they should have a significantly higher energy deposit yielding a consistent
tracking and calorimeter mass.

Each of the candidate tracks is projected to the front
face of the calorimeter.  If there is a local energy maximum in the 
calorimeter (peak tower) within one tower of the projected position,
we associate the charged particle track with that particular
calorimeter shower.  The calorimeter energy scale has been normalized such
that, on average, the sum of energy deposited in a three by three array
of towers surrounding and including the peak tower is equal to the 
charged particle's total kinetic energy. 
Thus, for each candidate we compute the total kinetic energy
from the nine tower sum.  

We require that the peak tower time-of-flight
agrees with the projected time from the hodoscopes within $\pm$ 2 ns.
If this agreement is met, since the hodoscope time resolution is significantly
better than the calorimeter resolution, we calculate the velocity 
$\beta$ using the three
hodoscopes only.  It should be noted that this calculation does not use
the assumed time at the target.
 
Before proceeding further, since real strangelets are expected to deposit
significantly more energy in the calorimeter than ``fake'' strangelets
(which are really protons striking the calorimeter), it is critical to eliminate
calorimeter showers which have energy contamination from other particles.   
Various cuts are applied to the calorimeter
information to remove contaminated energy showers.  Cuts are placed on the
time agreement of the eight side towers and the central position of the
cluster with respect to the position projected from tracking.  We also
require that there be no other energy peaks in the towers surrounding
the shower of interest.  The cuts maintain good
efficiency for ``real'' strangelet showers, while rejecting contaminated
ones.
The timing and energy cuts used for the
``+1.50T'' data set are slightly different, on account of differences in total
particle occupancy and overall background levels.  

\begin{figure}
\centerline{\hbox{\psfig{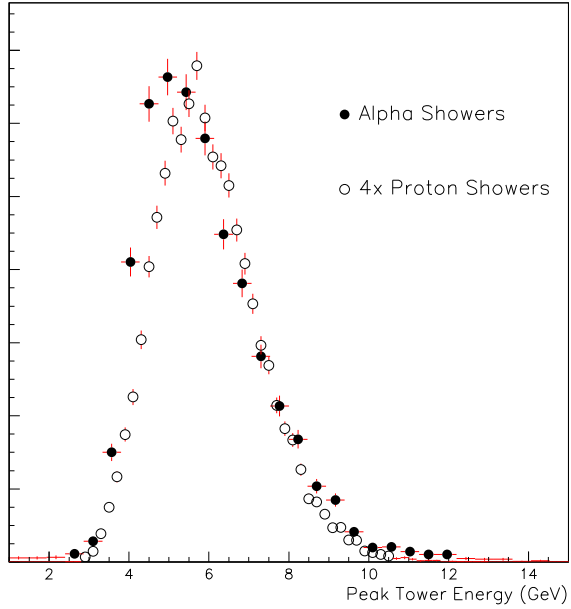}}}
\caption{The distribution of calorimeter measured energy (in the peak
tower only) for $^4$He particles.  These $^4$He have fired the high
mass trigger.  The $^4$He have a kinetic energy
of approximately 11 GeV, with ~50\% of the energy expected to
be deposited in the peak tower.
Also shown in open circles is the energy distribution
of four independent proton showers summed together with similar
kinetic energy per nucleon.}
\label{fig:fig_showers}
\end{figure}

In using the calorimeter, we assume that the strangelet has a
hadronic shower in the calorimeter.  We have assumed that light strangelets 
will fragment in their first one or two inelastic collisions into
their constituent baryons (and resulting mesons).  We have studied 
the calorimeter response to light nuclei such as $^4$He.  
Shown in Fig. \ref{fig:fig_showers} is the 
distribution of calorimeter energies (measured in the peak tower only)
from identified alpha particles.  
Also shown is the combination of four separate proton showers with
similar kinetic energy per nucleon to the alpha showers shown.  The
agreement is quite good and gives us confidence in our picture of the
interaction of multi-baryonic objects in the calorimeter.

\subsection{Mass versus Mass}

\begin{figure}
\centerline{\hbox{\psfig{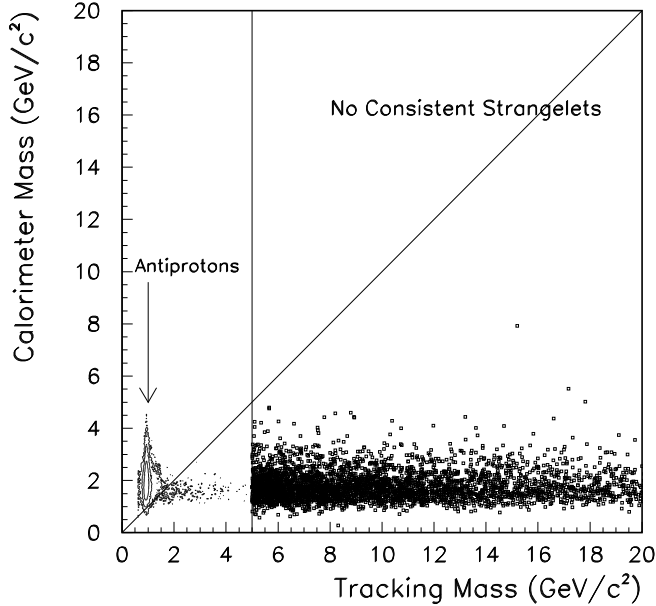}}}
\caption{The plot shows Z=$-$1 particle candidates measured masses from
tracking versus calorimeter reconstructed masses.  Shown as a contour plot
is the mass region from 0.0-5.0 GeV/c$^2$ with a clear peak of antiprotons.
All candidates with tracking mass greater than 5 GeV/c$^2$ are shown as
boxes.}
\label{fig:fig_cand_mass_cmass}
\end{figure}

For all candidates with associated 
uncontaminated calorimeter showers, we plot the tracking 
mass versus the calorimeter mass.
A distribution of the tracking mass versus the
calorimeter mass for all Z=$-$1 candidates is shown in
Fig. \ref{fig:fig_cand_mass_cmass}.  We have plotted the low mass
region as a contour plot to show the prominent antiproton
peak.  It is observed that the antiproton mass reconstructed in the
calorimeter using Eq. 2 is higher than the tracking mass.  This difference is expected
because antibaryons deposit not only their kinetic energy in the calorimeter,
but also twice the mass energy from the annihilation.  If we
recompute the antiproton mass from the calorimeter, removing the expected
annihilation energy, then the tracking and calorimeter masses are in good
agreement.  All candidates with a tracking mass above 5 GeV/c$^2$ are shown
as larger boxes.

In the mass range $5-10~\rm{GeV/c^{2}}$ the calorimeter mass is required to
agree with the tracking mass within $-1.0 \sigma$ and $+3.0 \sigma$,
where $\sigma$ is the $RMS$ mass resolution from the calorimeter.  This mass
resolution is dominated by the energy resolution of the calorimeter.   
There are no candidates which meet this requirement in this 
mass range.   

Above tracking mass of 10 GeV, the agreement requirement is loosened.
The calorimeter has been shown to have a linear energy response 
up to at least 12 GeV \cite{NIM},
but at some larger energy the photo-multiplier gains saturate and this linearity
is violated.  Additionally, the amplitude-to-digital converters, ADC's, have
a finite range only extending up to approximately 40 GeV per tower in 
measured energy.  Therefore high mass candidates which deposit a large
amount of energy may reconstruct a lower calorimeter mass than expected.  
Thus, we want to be careful about the calorimeter mass required for
high mass candidates.  There is only one potential candidate with tracking
mass greater than 10 GeV/c$^2$.  This candidate is shown in 
Fig. \ref{fig:fig_cand_mass_cmass} and has a tracking mass of 15.2~GeV/c$^2$ and
a calorimeter mass of 7.8~GeV/c$^2$.  The kinetic energy from
the tracking information is 61~GeV, while the calorimeter measured
energy is only 21~GeV.  The peak calorimeter
tower measured only 11.3~GeV, below the energy range where the gains
might saturate.  Therefore, this one potential candidate is 
considered background.

There have been interpretations of cosmic ray ``Centauro'' type events 
as resulting from heavy strangelets (A $>>$ 10) which penetrate through the
Earth's atmosphere \cite{Bjorken,Centauro}.  The strange quark droplets might
have a density significantly greater than normal nuclear matter density, 
and thus have a smaller inelastic cross section than a comparable nucleus.
However, any strangelet in our experiment strikes five hadronic interaction
lengths of lead (for a strangelet with the geometric size of a proton), and we
expect a negligible probability of significant energy leakage out the 
back of our calorimeter.

\begin{figure}
\centerline{\hbox{\psfig{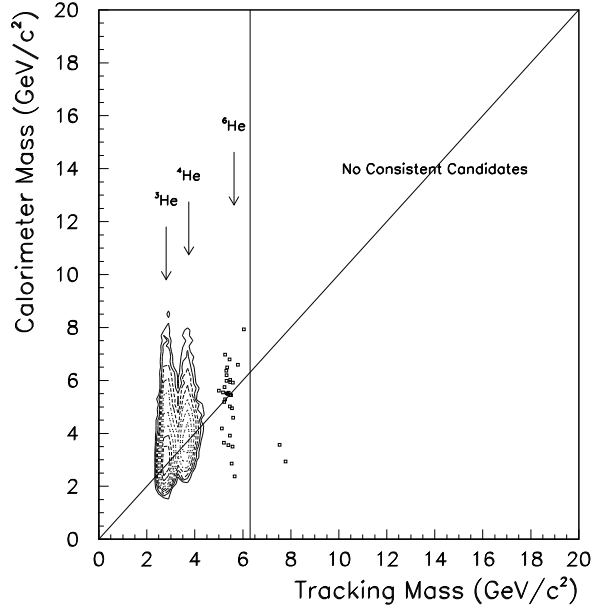}}}
\caption{The scatter plot is Z=+2 particle candidates measured mass
from tracking versus calorimeter reconstructed mass.  Shown as a contour
plot is the mass region from 0.0-5.0 GeV/c$^2$ with peaks from $^3$He and
$^4$He.  All candidates above this mass are shown as squares.  The only
particles in this mass range with consistent tracking and calorimeter
masses are identified as $^6$He isotopes.}
\label{fig:fig_pos_mass}
\end{figure}

Similar analysis for charge Z~=~+1 from the ``+1.50T'' data reveals no strangelet
candidates after consideration of the calorimeter and tracking information.
As stated before, there are no Z~=~$-$2 candidates, even before considering
the calorimeter response.  The distribution
of charge Z~=~+2 candidates from the ``+1.50T'' data sample is shown in 
Fig. \ref{fig:fig_pos_mass}.  The mass range below 5 GeV/c$^2$ is shown
as a contour plot where one can see clear peaks
for $^3$He and $^4$He isotopes.  All points above tracking mass of 5 GeV/c$^2$ 
are shown as boxes.  There are no consistent candidates
above A=6.  From the two field settings, there are approximately 50
$^6$He nuclei measured within $\pm$ 0.6 units of mid-rapidity.  These
nuclei are believed to be not from beam or target fragments, but rather
from the coalescence of separate nucleons \cite{Coal}.  This is the first
significant measurement of A=6 coalescence yields at these energies, which represents a
true six particle correlation.  We will discuss the importance of these
yields in the final section of this paper.

In summation, there are no remaining consistent candidates for Z=$\pm$1,$-$2 above
mass 5 GeV/c$^2$, and Z=+2 above mass 6 GeV/c$^2$.  Thus, we
use this information to set upper limits on the production of strangelet states.

\section{Production Limits}

In order to relate the null result to overall production limits, we need to
calculate the fraction of possible strangelets we would have measured.
The acceptances and efficiencies vary for different strangelet species and
depend on the expected kinematic distribution of strangelets.  

\begin{figure}
\centerline{\hbox{\psfig{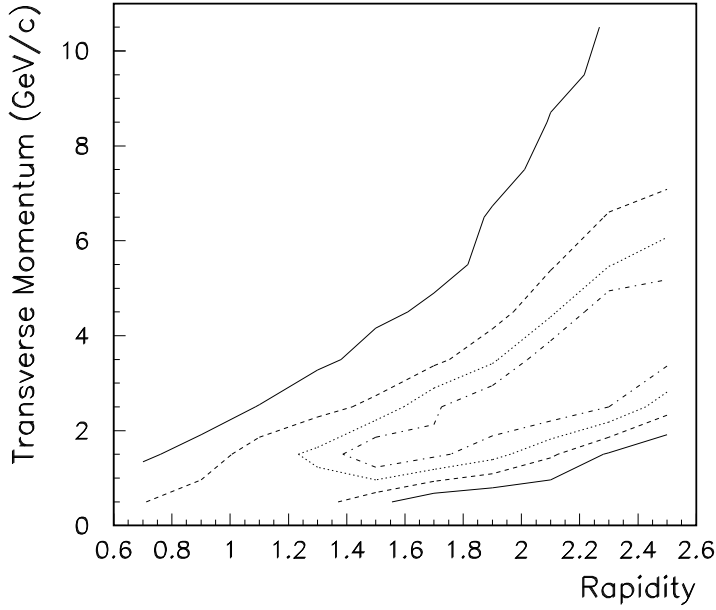}}}
\caption{The geometric acceptance for strangelets (A=20, Z=$-$1) 
at the ``$-$0.75T'' field setting is shown as
a function of rapidity and transverse momentum.  The inner
most and outer most contour lines surround acceptance values in excess of 25\%
and 3\%, respectively.}
\label{fig:fig_accept}
\end{figure}

Strangelets, since they are yet to be discovered, have an unknown momentum distribution.
We assume the following production model:
\begin{equation}
{{d^{2}N} \over {dydp_{t}}} \propto p_{t} e^{-{{2p_{t}} \over {<p_{t}>}}} e^{-{{(y-y_{NN})^{2}} \over {2\sigma_{y}^{2}}}}
\end{equation}
We further assume that the mean transverse momentum $<p_{t}> = 0.6 \space \sqrt{A}$~GeV/c, 
where A is the mass of the strangelet in baryon number, and
$\sigma_{y}$~=~0.5 is the width of a Gaussian rapidity 
distribution centered at mid-rapidity ($y_{NN}=1.6$).  
The rapidity and transverse momentum distributions are
assumed to be uncorrelated.  Shown in Fig. \ref{fig:fig_accept} is the
geometric acceptance for a charge one, A=20 strangelet as a function
of rapidity and transverse momentum.  The coverage extends over a wide
range of rapidity and $p_t$, thus making the experimental results 
relatively insensitive to the expected distribution.  It should be
noted that we do not have acceptance extending to $p_{t}=0$ due to
the physical constraint of the vacuum chamber.  For example, a mid-rapidity
A=20 strangelet must have at least 
800~MeV/c (40~MeV/c per nucleon) of transverse momentum to
be detected in the downstream spectrometer.

The final strangelet limits are quoted as 90\% confidence
level upper limits in 10\% most central Au+Pb interactions at 11.6~A~GeV/c.  
The limit is given as:
\begin{equation}
90\% \rm{C.L.}~=~{{N_{Poisson}} \over {N_{sampled}}}~~{{1} \over {\epsilon_{accept} \times \epsilon_{tracking} \times \epsilon_{calorimeter} \times \epsilon_{trigger}}}
\end{equation}
where the 90\% confidence level limit from Poisson statistics is $N_{Poisson} = 2.30$ and
$N_{sampled}$ is the total number of events sampled.  The various efficiencies $\epsilon$
are described below.

The efficiencies vary with strangelet species (A,S) and with the
production model, but typical values are given below.
The overall geometric acceptance $\epsilon_{accept}$ is approximately
8\%.  The tracking efficiency $\epsilon_{track}$ including 
track quality cuts is approximately 75\%.  
The calorimeter contamination cut efficiency $\epsilon_{calorimeter}$ varies 
over quite a range depending on the incident particle occupancy from 40-80\%.  
Finally, the trigger efficiency
$\epsilon_{trigger}$ values are quite high, in the range of 90-100\%.
We have calculated these efficiencies using a full GEANT simulation 
of the experiment including the 
magnets, vacuum chamber, detectors, etc.  We have used detector survey 
data as input for the various detector geometries.  We use this simulation
for the calculation of geometric acceptance and single particle
tracking efficiency.   We determine the efficiency of each detector
(for example due to small gaps between scintillator slats in the hodoscopes)
by using the data to find tracks without using a given detector and then checking for
a consistent hit in that detector.
In order to determine the multi-track efficiencies and calorimeter shower
cut efficiencies, we have taken Monte Carlo detector hit information 
(simulating the measured detector responses), overlayed these hits 
with real experimental data, and processed the results through
our tracking and shower analysis.

\begin{figure}
\centerline{\hbox{\psfig{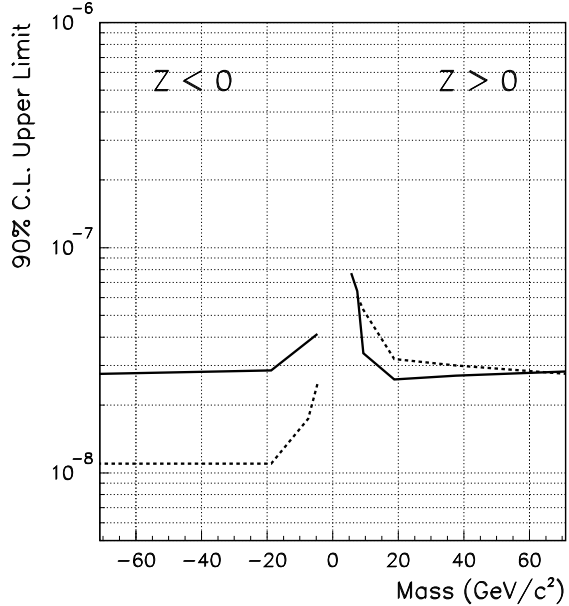}}}
\caption{Shown are the 90\% confidence level upper limits per 10\%
most central Au + Pb collision for negatively and positively
charged strangelet states as a function of strangelet mass.  The solid
lines are for Z=$\pm$1 and the dashed lines are for Z=$\pm$2.}
\label{fig:fig_limits}
\end{figure}

The upper limits for the two data sets combined are shown in 
Fig. \ref{fig:fig_limits}.  
The final limits do not depend significantly on the production model 
(at the level of a factor of two)
since the experiment is sensitive over a broad range of momentum space.  
The final limits are also relatively independent of the strangelet mass.

\begin{figure}
\centerline{\hbox{\psfig{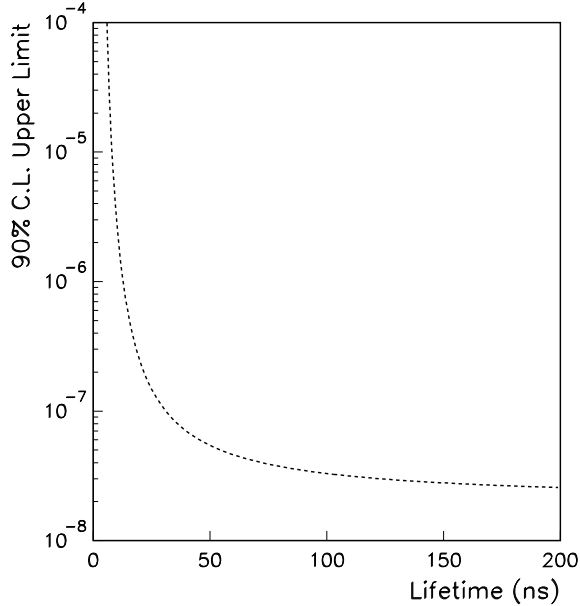}}}
\caption{Shown are the 90\% confidence level upper limits per 10\%
most central Au~+~Pb collision for negatively and positively
charged strangelet states as a function of strangelet lifetime.}
\label{fig:fig_str_lifetime}
\end{figure}

These limits are for metastable strangelets with proper lifetimes on the
order of $\tau \geq 50$~ns.  If the lifetime of the strangelet is less
than 50~ns, the sensitivity drops significantly.  Shown in
Fig. \ref{fig:fig_str_lifetime} is the lifetime dependence of
our upper limits.  The lifetime dependence is calculated using
a 100~ns flight time (approximately the time-of-flight to the
calorimeter), a relativistic factor $\gamma=2$, and an upper limit of
$2 \times 10^{-8}$ per central interaction for lifetimes significantly
greater than 50~ns.  

\section{Other Experimental Results}

There have been previous searches for charged strangelets in relativistic heavy
ion experiments.  To date no experiment has published results indicating 
a clear positive signal and so all have set production upper limits.  This statement
excludes $H^{0}$ dibaryon results for which there are experimental 
observations \cite{Longacre,E888},
but no definitive conclusions.  Earlier searches for charged strangelets
were performed using Si beams at 
the BNL-AGS by experiments E814 \cite{Rotondo} 
and E858 \cite{Aoki}, and using S and Pb beams at higher energies at the CERN-SPS 
by experiment NA52 \cite{Pretzl}.  All have published null results.  
The production potential (probability for strangelet formation) 
may be quite different for smaller colliding systems 
using Si and S beams or systems at significantly higher beam energy 
(which produce systems with a lower baryon density).  Therefore, 
it is difficult to compare these experimental limits with the present results.

There are two experiments at the BNL-AGS which have set limits in Au + Au (Pt)
collisions.  Both experiments E878 \cite{Beavis} and E886 \cite{Rusek}
set limits per minimum bias collision, which we need to relate to our
limits in the 10\% most central collisions.  If we assume that 50\% of all
strangelets are produced in the 10\% most central collisions (which
one might roughly expect given the coalescence calculations in \cite{Baltz}),
then our limits in central collisions should be divided by a factor
of five to convert to limits per minimum bias collision.  This results
in E864 limits at approximately $4 \times 10^{-9}$ per minimum bias collision.
Experiments E878 and E886 set limits in minimum bias collisions for negatively
charged strangelets at approximately $5 \times 10^{-9}$ and $1 \times 10^{-8}$
respectively, and for positively charged strangelets at 
approximately $5 \times 10^{-6}$ and $1 \times 10^{-7}$ respectively.  
%In all
%cases the E864 limits are lower.

It is critical to note that both E886 and E878 are focusing spectrometer 
experiments which operate with maximum rigidity settings of 2 GeV/c/Z and
20 GeV/c/Z, respectively.  Since the production of strangelets is expected to
be peaked at mid-rapidity (corresponding to a momentum per nucleon of 2 GeV/c),
these rigidity limitations make these experiments relatively insensitive to
strangelets with masses significantly greater than 10 GeV/c$^2$.  Thus, in 
the mass range 10-100 GeV/c$^2$, the limits set in this paper represent
by far the best limits to date.

\section{Constraining Production Models}

Ideally, we would like to use these limits to make a fundamental
statement about the stability of strange quark matter.  We would like
to constrain the bag model parameters to eliminate values which
predict metastable strangelets in this mass range.  
Even within the framework of the bag model, there is a correlated set
of parameters which have not been fully explored theoretically as
regards the strangelet states the model would predict.  Some regions
of this parameter space, of course, have been worked out and as noted
were the motivation for this experiment.  In addition, the mechanisms by
which strangelets would be produced in heavy ion collisions are not
well known and so one must consider different production models.  In the
analysis below, we examine the different production models and are
able to establish interesting and useful constraints, in some cases,
on the joint hypothesis regarding the models and particular
strangelet states.

%However, the
%predictions for strangelet stability in this mass range are not well
%determined.  Additionally, it is possible that charged
%strangelets are metastable, but that heavy ion collisions at these
%energies do not provide a sufficient environment in which to create
%strangelet states.

\subsection{Plasma models}

The most optimistic mechanism for the production of strangelets involves
the formation of a quark-gluon plasma \cite{Carsten}.  However, this scenario
is also the most difficult in which to produce reliable theoretical calculations
as to the exact rate of strangelet production.  If a quark-gluon plasma
is formed in some of the heavy ion collisions studied, the plasma is
expected to have a large net baryon density -- rich in $u$ and $d$ quarks 
and poor in $\overline{u}$ and $\overline{d}$ quarks.  This hot quark system cools
via the emission of mesons.  Since it is easier for antistrange
quarks to pair up with $u$ and $d$ quarks in $K^{+}$ ($u\overline{s}$) 
and $K^{0}$ ($d\overline{s}$), as opposed to strange
quarks pairing with $\overline{u}$ and $\overline{d}$ quarks, the hot plasma
gains net strangeness.  As the plasma cools it may form a droplet of
cold strange quark matter which might be metastable and be measurable
in our experiment.  

Our experiment cannot explicitly rule out any one of the steps
required in this process:  (1) QGP formation, (2) strangeness distillation,
and (3) metastable strangelet states.  However, we can state that in
approximately $10^{8}$ central collisions, the combination
of these three steps does {\bf not} occur.
\begin{equation}
\rm{Probability(1) \times Probability(2) \times Probability(3) < 10^{-8}}
\end{equation}
For example, if one believed that a QGP is formed in 1 out of 10,000 central Au+Pb
collisions, then the probability of this QGP forming a metastable
charged strangelet is less than 0.01\%.

There are specific predictions on production rates of
strangelets assuming a QGP phase transition.  Crawford {\it et al.}
calculate negatively and positively charged strangelet 
production levels in heavy ion collisions \cite{Crawford}.  
The model assumes that a quark-gluon plasma is formed in every 
10\% most central collision and determines how often a strangelet of 
given mass and charge is distilled out.  Their predictions for charged
strangelets of mass 10-20 GeV/c$^2$ from minimum bias Si+Au collisions are in the range
of $4 \times 10^{-11}$ to $8 \times 10^{-8}$.  One might expect higher
production rates in central Au + Pb collisions.  Our limits eliminate
some of the predictions; however, the calculations have many rough
assumptions and should not be viewed as exact within orders of magnitude.

\subsection{Coalescence Models}

A very different production mechanism for strangelets is via the
coalescence of strange and non-strange baryons.  In the QGP scenario,
a large fraction of the colliding system may form a bubble of hot
quark matter which may cool into a strangelet.  Thus, the strangelet could
have quite a large mass (A $>$ 15).  In a coalescence picture, after
the colliding system has expanded significantly, interactions between
particles become less frequent until eventually the particles are
free streaming (freeze-out).  At the point of freeze-out baryons which are
close to each other in configuration space and momentum may fuse together to
form nuclei or possibly hypernuclei.  Most hypernuclei are expected to have
lifetimes on the order of the $\Lambda$ particle and thus would decay before
traversing our spectrometer.   However if a strangelet state of similar
quantum numbers (A, S) were more stable than the hypernucleus, the nuclear
state might act as a doorway to the strange quark matter state.  

One can make relatively reliable coalescence rate calculations, which can
be checked with nuclear isotope yields.  In the paper of 
Baltz ${\it et~al.}$, the authors predict a rate of 
$^{7}_{\Xi^{0}\Lambda\Lambda}{\rm{He}}$ at
3-7$\times 10^{-8}$ per central Au + Au collision \cite{Baltz}, 
while our sensitivity for a strangelet state with
the same quantum numbers is $6 \times 10^{-8}$.  One might thus conclude
that we are testing limits within the coalescence model at the level of 
A~=~7 and $|S|=4$.  
However, recent results from our experiment for measured light
nuclei indicate that the calculations in \cite{Baltz} overestimate the experimentally
measured yield of light nuclear isotopes like $^3$He and $^4$He \cite{HIPAGS}.
Also, in the data sets discussed in this paper, we observe a significant number
of $^6$He isotopes, but no $^8$He states (note that $^7$He is unstable on
the lifetime scale required to be measured in our experiment).  Thus, we have
roughly reached a coalescence sensitivity of A=7 and $|S|=0$.  For each coalesced
baryon, the penalty for changing a non-strange baryon to a strange baryon is
estimated to be approximately 0.2 \cite{Baltz}.  Therefore, the experiment is 
roughly sensitive to states with 
\begin{equation}
|A| + 0.5 \times |S|~~~ < ~~~ 7
\end{equation}
We are beginning to address the coalescence production of strangelets for
relatively light states.  If we had observed a heavy strangelet state ($A>10$),
it would have been good evidence for the plasma distillation mechanism, since
this mass range is beyond what one would expect from coalescence. 

\section{Conclusions}
In experiment 864, using data taken in the fall of 1995, 
we have sampled nearly six billion central Au + Pb
collisions at the BNL-AGS.  Through the use of significant redundant
tracking measurements and calorimetry, we find no consistent 
candidates for new states of strange
quark matter.  This represents the lowest and most significant limit
to date in relativistic heavy ion collisions at these energies.
From data taken in the winter of 1997 and to be taken in 1998, we
will either extend these limits by approximately an order of
magnitude or possibly discover strangelets.  

%It is not possible to put firm constraints on strangelet stability, and
%difficult to completely rule out their formation in these collisions
%due to the inexactness of the theoretical calculations.  However, we
%now know that not one such state is produced in one hundred million
%central collisions.
  
\section{Acknowledgments}
We acknowledge the efforts of the AGS and Tandem staff in providing
the Au beam for the experiment.
This work was supported by grants from the Department of Energy (DOE)
High Energy Physics Division, DOE Nuclear Division, the
National Science Foundation, and the Istituto Nationale di 
Fisica Nucleare of Italy (INFN).

Finally, we would like to record here the respect, affection, and
appreciation which our collaboration shares for our late colleague, Carl Dover.
It should be noted that this experiment, E-864 at the BNL AGS, owed
a great deal to Carl's insight and knowledge.  He was instrumental in
both the original conception of the experiment and in its design and scope.
His wisdom, and his friendship will be sorely missed.

%\begin{references}

\end{document}